\documentclass[12pt]{article}
\usepackage{a4,amsmath,cite,amssymb,graphicx,epsfig}

\def\ecart{\noalign{\medskip}}

\begin{document}

\begin{titlepage}

\vspace*{1cm}

\begin{center} 
\setlength{\baselineskip}{24pt}
{\LARGE THE ODDERON AT RHIC AND LHC}
\end{center}

\begin{center}
\vspace{2cm}
{\large Basarab Nicolescu}

\vspace{0.5cm} 
 Theory Group,  Laboratoire de Physique Nucl\'eaire  et des Hautes  \'Energies 
(LPNHE)\footnote{Unit\'e  de Recherche des Universit\'es 
  Paris 6 et Paris 7, Associ\'ee au CNRS},
 CNRS and Universit\'e Pierre et Marie Curie, Paris\\ 
 
 \vspace{1cm}
{\large \textbf{To the memory of Leszek Lukaszuk (1938 - 2007)}}

\end{center}

\vspace{2cm}


%
\begin{abstract}
The Odderon remains an elusive object, 33 years after its invention. 
The Odderon is now a fundamental object in QCD and CGC and it has to be found experimentally if QCD and CGC are right.  
In the present talk, we show how to find it at RHIC and LHC. 
The most spectacular signature of the Odderon is the predicted difference between the differential cross-sections for proton-proton and antiproton-proton at high $s$ and moderate $t$. 
This experiment can be done by using the STAR detector at RHIC and by combining these future data with the already present UA4/2 data.  
The Odderon could also be found by ATLAS experiment at LHC by performing a high-precision measurement of the real part of the hadron elastic scattering amplitude at small $t$.
\end{abstract}

\end{titlepage}
\newpage

\section{Introduction}
This contribution to EDS07 is based upon  work done in collaboration with Regina F. Avila and Pierre Gauron~\cite{Avila:2006wy}.

The Odderon is defined as a singularity in the complex J-plane, located at $J=1$ when $t=0$ and which contributes to the odd-under-crossing amplitude $F_-$.
The concept of Odderon first emerged in 1973 in the context of asymptotic theorems~\cite{Lukaszuk:1973nt}.
7 years later, it was possibly connected with 3-gluon exchanges in perturbative QCD~\cite{Bartels:1980pe,Jaroszewicz:1980mq,Kwiecinski:1980wb}, but it took 27 years to firmly rediscover it in the context of pQCD~\cite{Bartels:1999yt}.
The Odderon was also rediscovered recently in the Color Glass Condensate (CGC) approach~\cite{Hatta:2005as,Jeon:2005cf} and in the dipole picture~\cite{Kovchegov:2003dm}.
One can therefore assert that the Odderon is a crucial test of QCD.

On experimental level, there is a strong evidence for the non-perturbative Odderon: the discovery, in 1985, of a difference between $(d\sigma/dt)_{\bar pp}$ and 
$(d\sigma/dt)_{pp}$ in the dip-shoulder region $1.1<\vert t\vert <1.5$ GeV$^2$ at $\sqrt{s}=$ 52.8 GeV~\cite{Breakstone:1985pe,Erhan:1984mv}.
Unfortunately, these data were obtained in one week, just before ISR was closed and therefore the evidence, even if it is strong (99,9 \% confidence level), is not totally convincing.

 The maximal Odderon~\cite{Lukaszuk:1973nt,Nicolescu:2005bn}, is a special case (tripole) corresponding to the maximal asymptotic $(s\to\infty)$ behavior allowed by the general principles of strong interactions:
 \begin{equation}
\label{eq:1}
\sigma_T(s)\propto \ln^2s,\quad\mbox{as } s\to\infty
\end{equation}
and
\begin{equation}
\label{eq:2}
\Delta\sigma(s)\equiv \sigma_T^{\bar pp}(s)-\sigma_T^{pp}(s)\propto \ln s,\quad\mbox{as } s\to\infty\ .
\end{equation}
Interestingly enough, an important stream of theoretical papers concern precisely the maximal behavior~\cite{Lukaszuk:1973nt}, which was first discovered by Heisenberg in 1952~\cite{Heisenberg:1952zp} and later proved, in a more rigorous way by Froissart and Martin~\cite{Froissart:1961ux,Martin:1965jj}.
Half a century after the discovery of Heisenberg, this maximal behavior~(\ref{eq:1}) was also proved in the context of the AdS/CFT dual string-gravity theory~\cite{Giddings:2002cd} and of the Color Glass Condensate approach~\cite{Ferreiro:2002kv}.
It was also shown to provide the best description of the present experimental data on total cross-sections~\cite{Cudell:2001pn,Eidelman:2004wy}.

The maximal behavior of $ImF_+(s,t=0)\propto\ln^2s$ is naturally associated with the maximal behavior $ImF_-(s,t=0)\propto\ln s$.
In other words, strong interactions should be as strong as possible.

\section{Strategy}
In the present paper we will consider a very general form of the hadron amplitudes compatible with both the maximal behavior of strong interaction at asymptotic energies and with the well established Regge behavior at moderate energies, i.e. at pre-ISR and ISR energies~\cite{Gauron:1986nk,Gauron:1989cs}.

Our strategy is the following:
\begin{itemize}
  \item [1.]
  We will consider two cases: one in which the Odderon is absent and one in which the Odderon is present.
  \item [2.]
  We will use the two respective forms in order to describe the 832 experimental points for $pp$ and $\bar pp$ scattering, from PDG Tables, for $\sigma_T(s),\ \rho(s)$ and 
  $d\sigma/dt(s,t)$, in the s-range
  \begin{equation}
\label{eq:3}
4.539\mbox{ GeV }\leqslant\sqrt{s}\leqslant 1800\mbox{ GeV}
\end{equation}
and in the t-range
\begin{equation}
\label{eq:4}
0\leqslant \vert t\vert \leqslant 2.6\mbox{ GeV}^2\ .
\end{equation}
The best form will be chosen.
  \item [3.]
  In order to make predictions at RHIC and LHC energies, we will insist on the best possible \textit{quantitative} description of the data.
  \item [4.] 
 From the study of the interference between $F_+(s,t)$ and  $F_-(s,t)$ amplitudes we will conclude which are the best experiments to be done in order to detect in a clear way the Odderon.
\end{itemize}

\section{The form of the amplitudes}
$F_\pm$ are defined to be
\begin{equation}
\label{eq:5}
F_\pm(s,t)=\frac{1}{2}\left(F_{pp}(s,t)\pm F_{\bar pp}(s,t)\right)
\end{equation}
and are normalized so that
\begin{equation}
\label{eq:6}
\sigma_T(s)=\frac{1}{s}\mathrm{Im} F(s,0)\ ,\quad
\rho(s)=\frac{\mathrm{Re}F(s,t=0)}{\mathrm{Im}F(s,t=0)}
\end{equation}
\begin{equation}
\label{eq:7}
\frac{d\sigma}{dt}(s,t)=\frac{1}{16\pi s^2}\vert F(s,t)\vert^2\ .
\end{equation}
The $F_+(s,t)$ amplitude is written as a sum of the Regge poles and cuts in standard form~\cite{Avila:2006wy} and the Heisenberg component  $F_+^H(s,t)$ representing the contribution of a 3/2 - cut collapsing, at $t=0$, to a triple pole located at $J=1$ and which satisfies the Auberson-Kinoshita-Martin asymptotic theorem~\cite{Auberson:1971ru}:
\begin{equation}
\label{eq:8}
\begin{array}{rcl}
     \dfrac{1}{is} F_+^{H}(s,t) & = & H_1\ln^2\bar s\ \frac{2J_1(K_+\bar\tau)}{K_+\bar\tau}
\exp(b_1^+t) \\
\ecart
      &   + & H_2\ln\bar sJ_0(K_+\bar\tau)\exp(b_2^+t)\\
\ecart
      &   + & H_3[J_0(K_+\bar\tau)-K_+\bar\tau J_1(K_+\bar\tau)]\exp(b_3^+t)\ ,
\end{array}
\end{equation}  
where $J_n$ are Bessel functions, $H_k,\ b_k^+(k=1,2,3)$ and $K_+$ are constants,
\begin{equation}
\label{eq:9}
\bar s=\left(\frac{s}{s_0}\right)\exp\left(-\frac{1}{2}i\pi\right),
\mbox{ with }s_0=1\mbox{ GeV}^2	
\end{equation}
and
\begin{equation}
\label{eq:10}
\bar\tau=\left(-\frac{t}{t_0}\right)^{1/2}\ln\bar s,\quad \mbox{with }t_0=1\mbox{ GeV}^2\ .
\end{equation}

In its turn, the $F_-(s,t)$ amplitude is written as a sum of the Regge poles and cuts in standard form~\cite{Avila:2006wy} and $F_-^{MO}(s,t)$ representing the maximal Odderon contribution, resulting from two complex conjugate poles collapsing, at $t=0$, to a dipole located at $J=1$ and which satisfies the Auberson-Kinoshita-Martin asymptotic theorem:
\begin{equation}
\label{eq:24}
  \frac{1}{s}F_-^{MO}(s,t)=O_1\ln^2\bar s\frac{\sin(K_-\bar\tau)}{K_-\bar\tau}\exp(b_1^-t)+
O_2\ln\bar s\cos(K_-\bar\tau)\exp(b_2^-t)+O_3\exp(b_3^-t)\ ,
\end{equation}
where $O_k,\ b_k^-(k=1,2,3)$ and $K_-$ are constants.

\section{Numerical results}
Let us first consider the case \textit{without the Odderon}.
In this case, one has 23 free parameters.

In spite of the quite impressive number of free parameters, the $\chi^2$-value is inacceptably bad:
\begin{equation}
\label{eq:42}
\chi^2/dof=14.2\ .
\end{equation}
A closer examination of the results reveals however an interesting fact: the no-Odderon case describes nicely the data in the t-region $0\leqslant\vert t\vert\leqslant 0.6 \mbox{ GeV}^2$ but totally fails to describe the data for higher t-values.

 This failure does not mean the failure of the Regge model, which is a basic ingredient of the approach presented in this paper.
 It simply means the need for the Odderon.
 
 In the case with the Odderon, we have 12 supplementary free parameters.
 
 The total of 35 free parameters of our approach could be considered, at a superficial glance, as too big.
However, one has to realize that the 23 free parameters associated with the dominant $F_+(s,t)$ amplitude and with the component of $F_-(s,t)$ responsible for describing the data for $\Delta\sigma(s)$ (see eq.~(\ref{eq:2})) and $\Delta\rho(s,t=0)$, where
\begin{equation}
\label{eq:44}
\Delta\rho(s,t=0)\equiv\rho^{\bar pp}(s,t=0)-\rho^{pp}(s,t=0)
\end{equation}
are, almost all of them, well constrained.

Moreover, the discrepancy between he no-Odderon model and the experimental data in the moderate-t region (especially at $\sqrt{s}=52.8$ GeV and $\sqrt{s}=541$ GeV) is so big that, in their turn, the supplementary 12 free parameters (at least, most of them) are also well constrained.

Let us also note that the above - mentioned discrepancy in the region of $t$ defined by
\begin{equation}
\label{eq:45}
0.6< \vert t\vert \leqslant 2.6 \mbox{ GeV}^2
\end{equation}
can not come, as one could thing, from the contributions induced by  perturbative QCD.
The region (\ref{eq:45}) is fully in the domain of validity of the non-perturbative Regge pole model and the respective values of $t$ are too small in order to make pQCD calculations.

The resulting value of $\chi^2$ is
\begin{equation}
\label{eq:46}
\chi^2_{dof}=2.46\ ,
\end{equation}
an excellent value if we consider the fact that we did not take into account the systematic errors of the experimental data.

The partial value of $\chi^2$, corresponding only to $t=0$ ($\sigma_T$ and $\rho$) data is 
\begin{equation}
\label{eq:47}
\left.\chi^2_{dof}\right\vert_{t=0}=1.42\ ,
\end{equation}
an acceptable value (276 experimental forward points took into account).
Of course, better $\chi^2$ values can be obtained in fitting \textit{only} the $t=0$ data, as it is in often made in phenomenological papers.
However, it is obvious that, in a global fit including non-forward data, the corresponding $t=0$ parameters will be modified and therefore a higher $\chi^2$ value will be obtained.
The $t=0$ and $t\neq 0$ data are certainly independent but the parameter values are obviously correlated in a global fit.

\section{Predictions}
We plot in Fig. 1 our fit and predictions for $d\sigma/dt$ data at $\sqrt{s}=52.8$ GeV, at the RHIC energy values $\sqrt{s}=500$ GeV, at the commissioning run energy value $\sqrt{s}=900$ GeV and at the LHC energy value $\sqrt{s}=14$ TeV.
The description of the data at $\sqrt{s}=52.8$ GeV as offered by our approach is the best one existing in literature.
It has to ne noted that the structure (dip) region moves slowly, with increasing energy, from $\vert t\vert\approx 1.35$ GeV$^2$ at $\sqrt{s}=52.8$ GeV towards
$\vert t\vert \simeq 0.35$ GeV$^2$ at $\sqrt{s}=14$ TeV.

There is an interesting phenomenon of oscillations present in $\Delta(\frac{d\sigma}{dt})$ (see Fig. 2), where
\begin{equation}
\label{eq:48}
\Delta\left(
\frac{d\sigma}{dt}\right)(s,t)\equiv
\left\vert
\left(\frac{d\sigma}{dt}\right)^{\bar pp}(s,t)-
\left(\frac{d\sigma}{dt}\right)^{pp}(s,t)
\right\vert\ ,
\end{equation}


\noindent due of the oscillations present in the Heisenberg-type amplitude $F_+^H(s,t)$ and in the maximal Odderon amplitude $F_-^{MO}(s,t)$.
Unfortunately, we can not directly test the existence of these oscillations at RHIC and LHC energies, simply because we will not have both $pp$ and $\bar pp$ accelerators at these energies.
However a chance to detect these oscillations at the RHIC energy $\sqrt{s}=500$ GeV still exists, simply because the UA4/2 Collaboration already performed a high-precision $\bar pp$ experiment at a very close energy - 541 GeV~\cite{Augier:1993sz}.
By performing a very precise experiment at the RHIC energy $\sqrt{s}=500$ GeV and by combining the corresponding $pp$ data with the UA4/2 $\bar pp$ high-precision data one has a non-negligible chance to detect an oscillation centered around $\vert t\vert\simeq 0.9$ GeV$^2$ and therefore to detect the Odderon.
It is precisely the oscillation centered around $\vert t\vert\simeq 0.9$ GeV$^2$
 which is the reminder of the already seen oscillation centered around $\vert t\vert\simeq 1.35$ GeV$^2$ at the ISR energy $\sqrt{s}=52.8$ GeV.
  
 The participants at the workshop "Odderon Searches at RHIC", hold at BNL in September 2005, concluded that the best available setup for the experimental search for the Odderon is the proposed combination of STAR experiment and Roman pots at $pp2pp$ experiment,  described in the proposal "Physics with Tagged Forward Protons with the STAR detector at RHIC".
They also concluded that the most unambiguous signature of the Odderon is to detect a non-zero difference between $pp$ and $\bar pp$ differential cross-sections  at $\sqrt{s}=500$ GeV, as described above.
RHIC is an ideal place for discovering the Odderon and therefore testing QCD and CGC~\cite{Guryn:yk}.

LHC is also a good place to discover the Odderon.
We predict
\begin{alignat}{1}
&\sigma_T^{pp}(\sqrt{s}=14 \mbox{ TeV})=123.32\mbox{ mb}\ ,\label{eq:55}\\
&\Delta\sigma(\sqrt{s}=14 \mbox{ TeV})=-3.92\mbox{ mb}\ ,\label{eq:56}\\
&\rho_{pp}(\sqrt{s}=14 \mbox{ TeV},\  t=0)=0.103\label{eq:57}\ ,
\end{alignat}
and
\begin{equation}
\label{eq:58}
\Delta\rho(\sqrt{s}=14 \mbox{ TeV},\ t=0)=0.094\ .
\end{equation}
A $\rho^{pp}$-measurement at LHC would be certainly a very important test of the maximal Odderon, given the fact that our prediction is sufficiently lower than what dispersion relations without Odderon contributions could predict ($\rho\simeq 0.12 - 0.14$).

There are several other proposals for detecting the Odderon, summarized in the nice review written by Ewerz~\cite{Ewerz:2003xi}.

\section{Conclusions}
There are very rare cases in the history of physics that a scientific and testable idea is neither proved nor disproved 33 years after its invention.
The Odderon remains an elusive object in spite of intensive research for its experimental evidence.

The main reason for this apparent puzzle is that most of the efforts were concentrated in the study of $pp$ and $\bar pp$ scattering, where the $F_-(s,t)$ amplitude is hidden by the overwhelming $F_+(s,t)$ amplitude.
The most spectacular signature of the Odderon is the predicted difference between $pp$ and $\bar pp$ scattering at high $s$ and relatively small $t$.
However, it happens that, after the closure of ISR, which offered the first strong hint for the existence of the Odderon, there is no place in the world where $pp$ and $\bar pp$ scattering are or will be measured at the same time. 
This is the main reason of the non-observation till now of the Odderon.

We show that we can escape from this unpleasant situation by performing a high-precision measurement of $d\sigma/dt$ at RHIC, at $\sqrt{s}= 500$ GeV, and by combining these future data with the already present high-precision UA4/2 data at $\sqrt{s}= 541$ GeV.

There is no doubt about the theoretical evidence for the Odderon both in QCD and CGC.
The Odderon is a fundamental object of these two approaches and it has to be found at RHIC and LHC if QCD and CGC are right.

\section*{Acknowledgments}
I am grateful to Professors Claude Bourrely and Evgenij Martynov who kindly repeated our calculations. 
This crosscheck proved to be very useful.

I dedicate this talk to the memory of Leszek Lukaszuk (1938-2007), who was not only a brilliant physicist but also an extraordinary human being and an incomparable friend.


\newpage
\begin{figure}[h]
\includegraphics[scale=0.4]{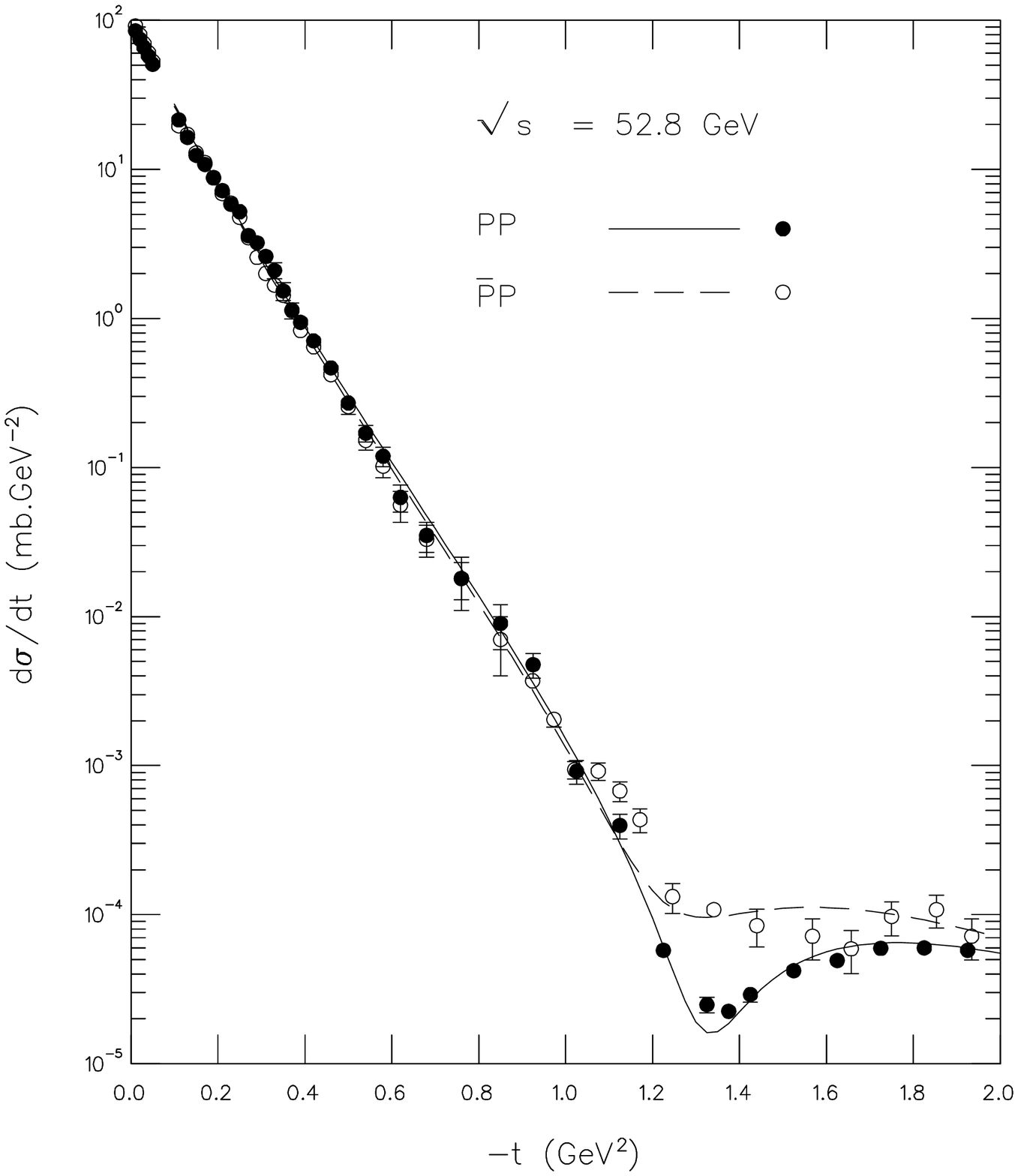}\hspace{0.5cm}\includegraphics[scale=0.4]{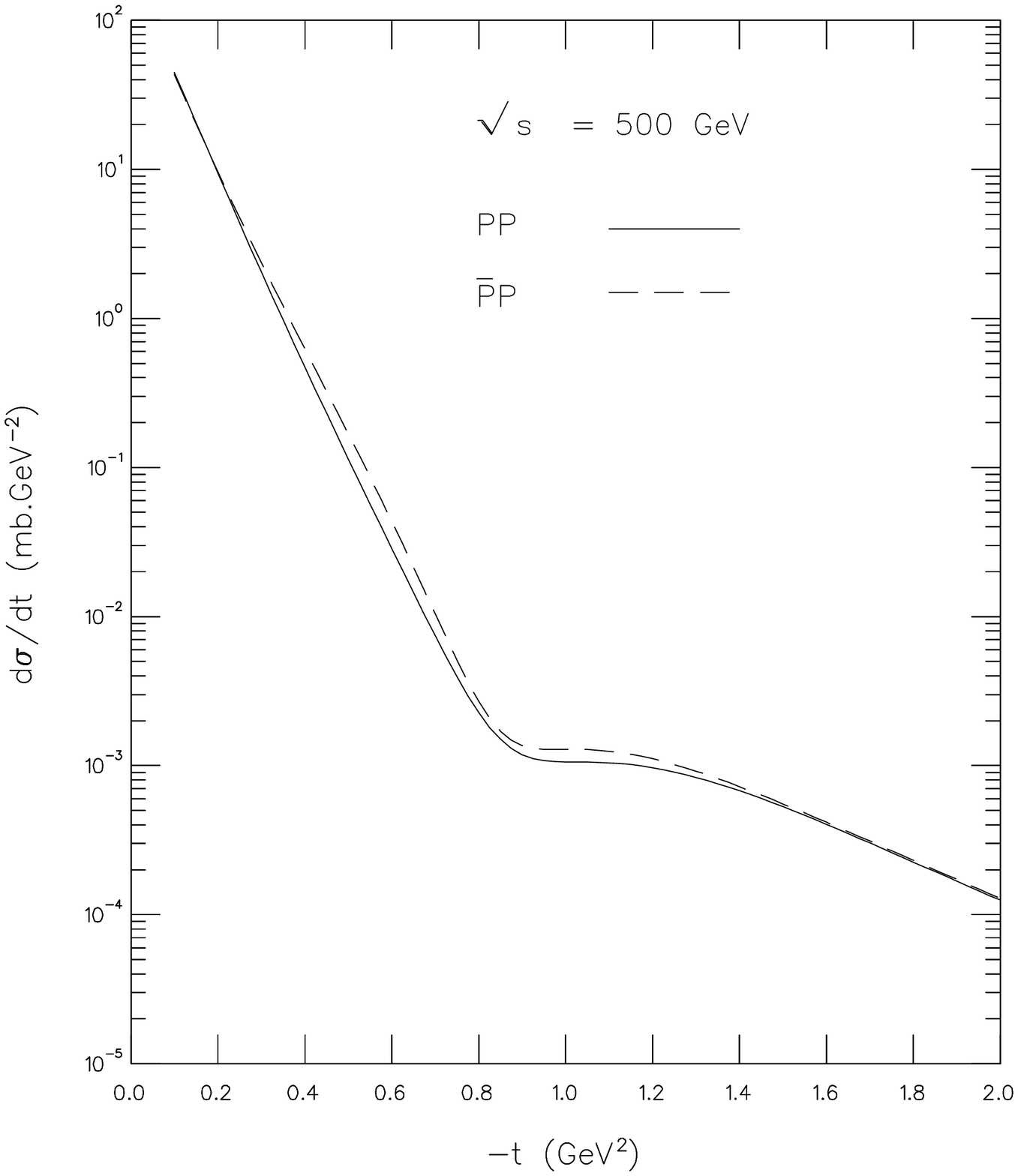}\\
\includegraphics[scale=0.4]{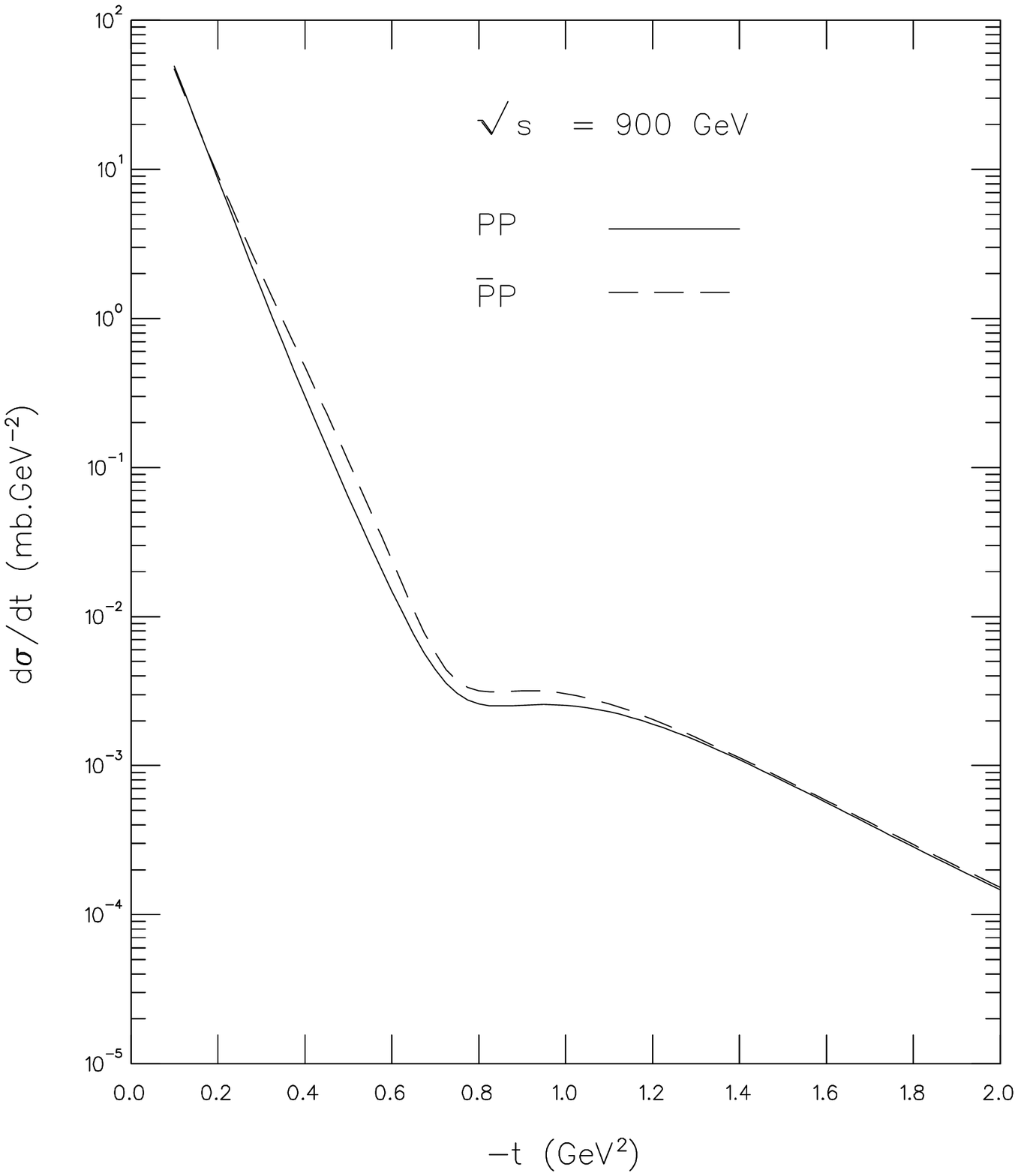}\hspace{0.5cm}\includegraphics[scale=0.4]{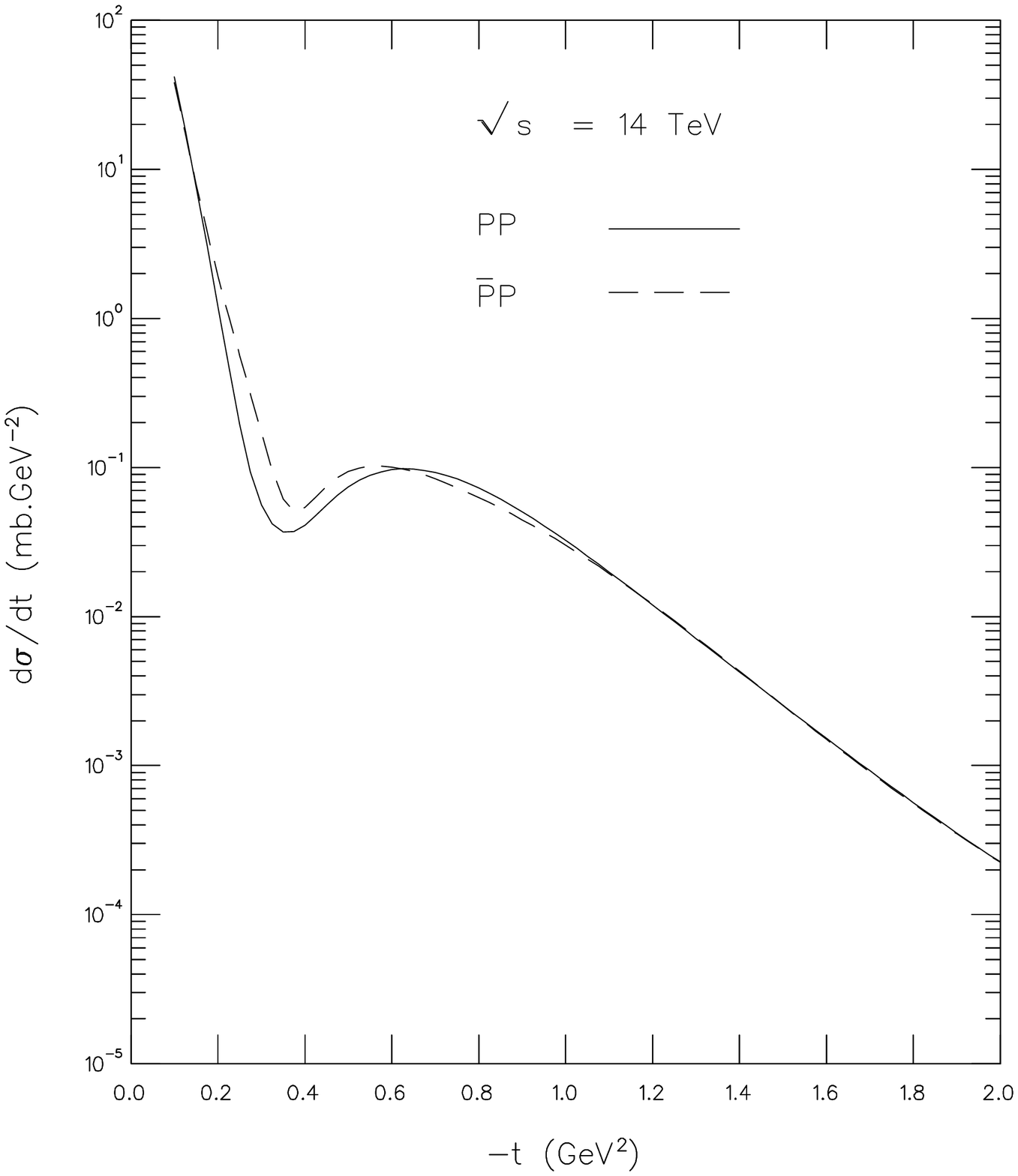}
\caption{The structure (dip) region moves slowly, with increasing energy, from $\vert t\vert\approx 1.35$ GeV$^2$ at $\sqrt{s}=52.8$ GeV towards
$\vert t\vert \simeq 0.35$ GeV$^2$ at $\sqrt{s}=14$ TeV.}
\end{figure}

\begin{figure}[h]
\includegraphics[scale=0.4]{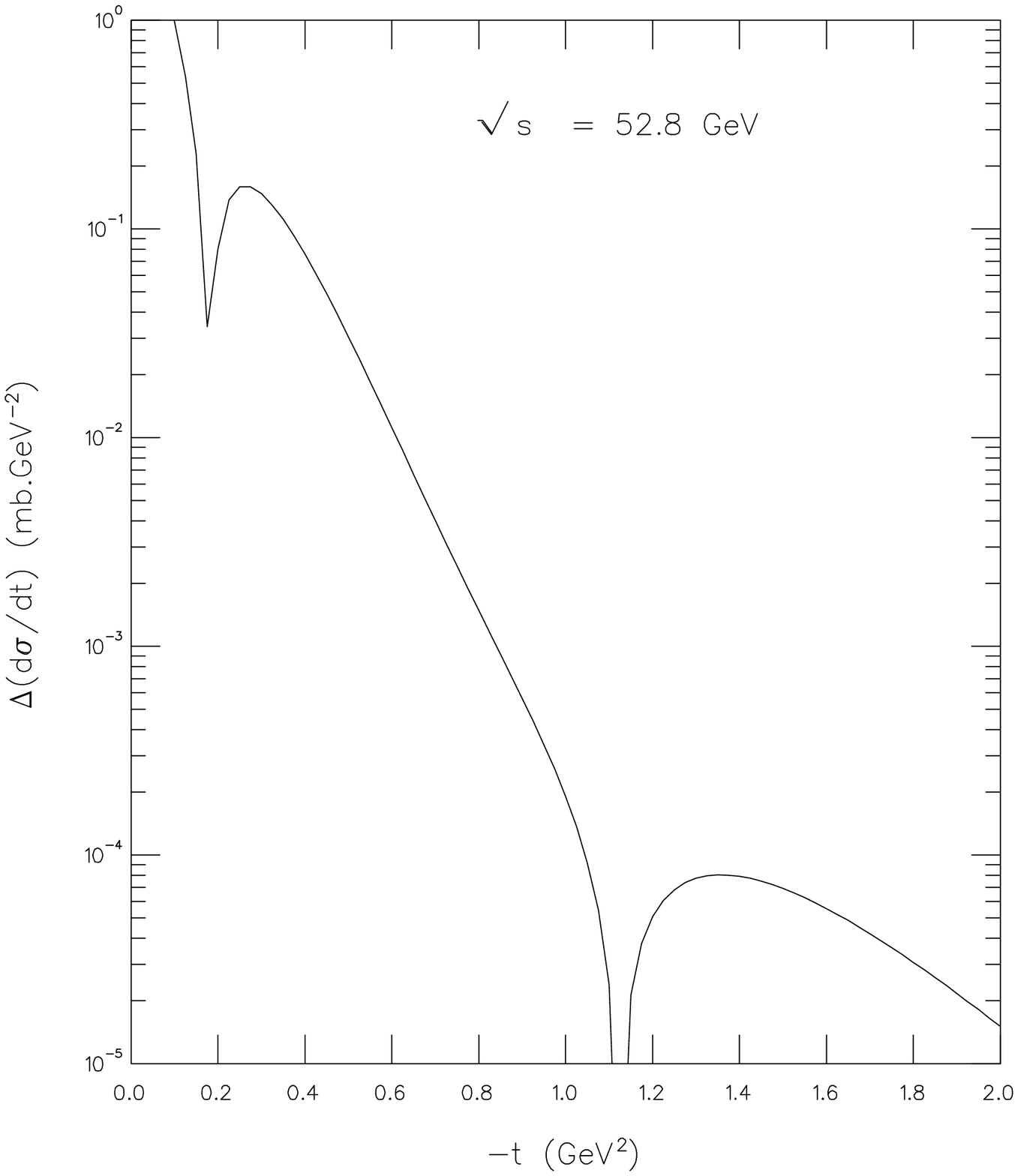}\hspace{0.5cm}\includegraphics[scale=0.4]{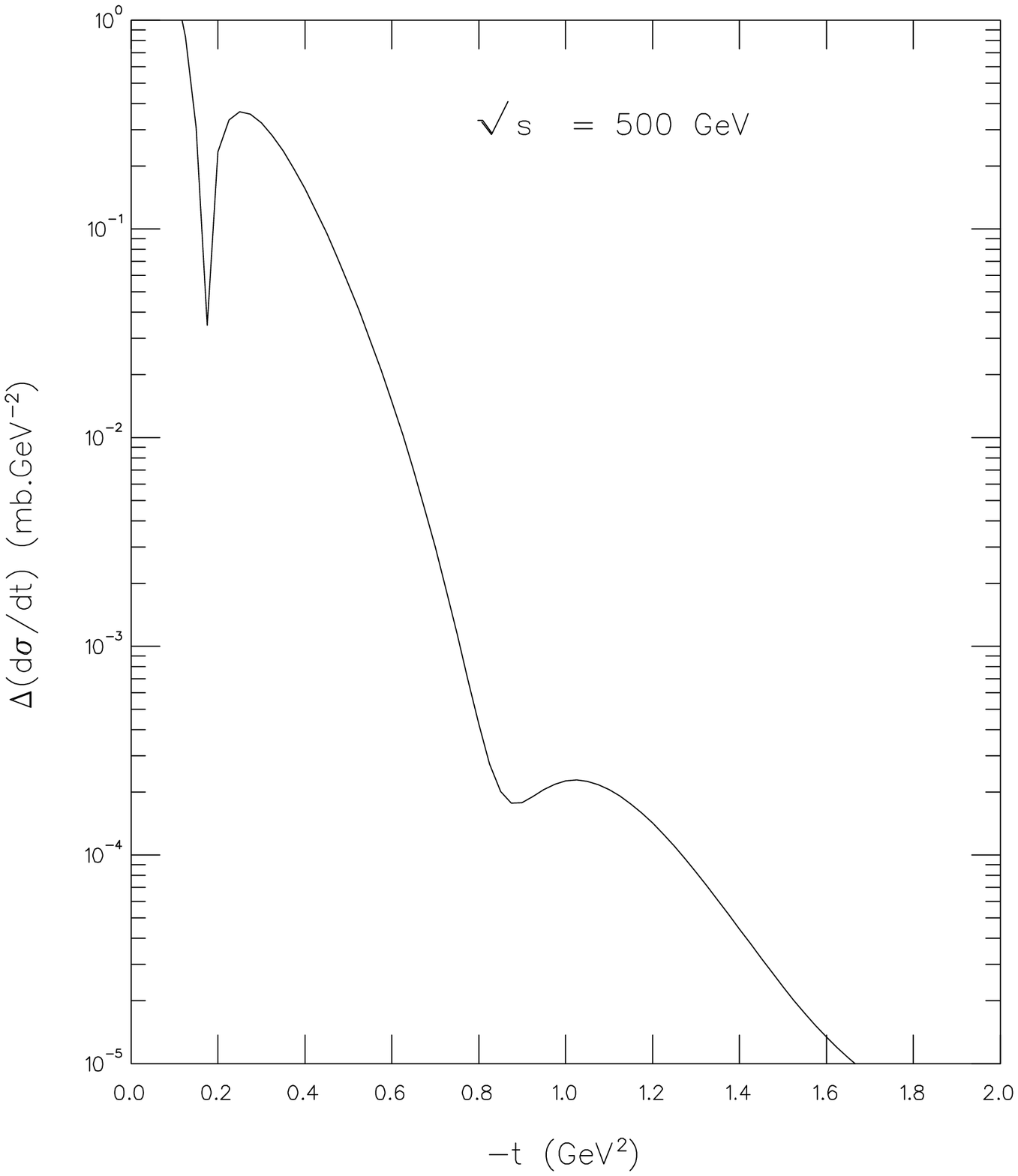}\\
\includegraphics[scale=0.4]{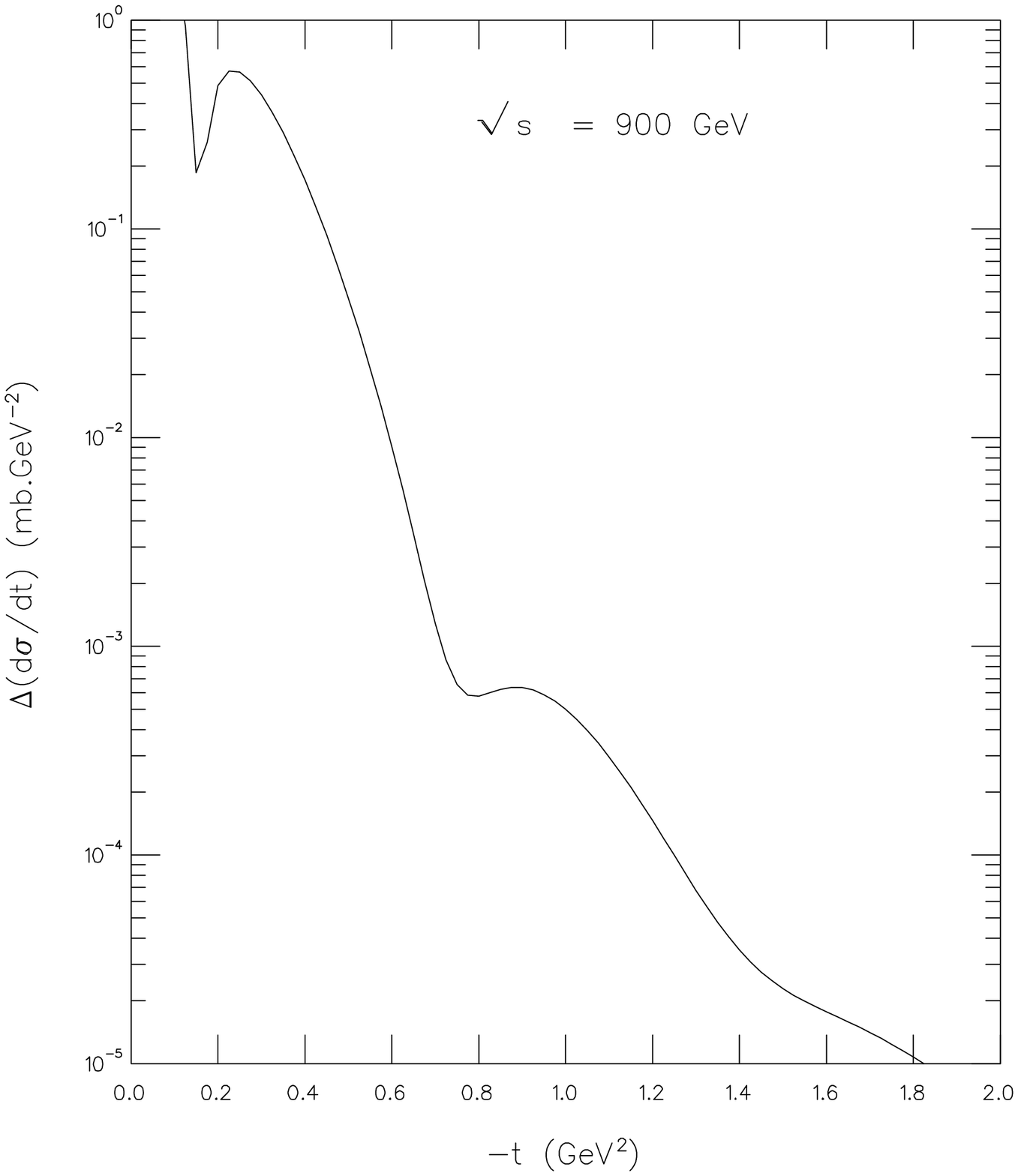}\hspace{0.5cm}\includegraphics[scale=0.4]{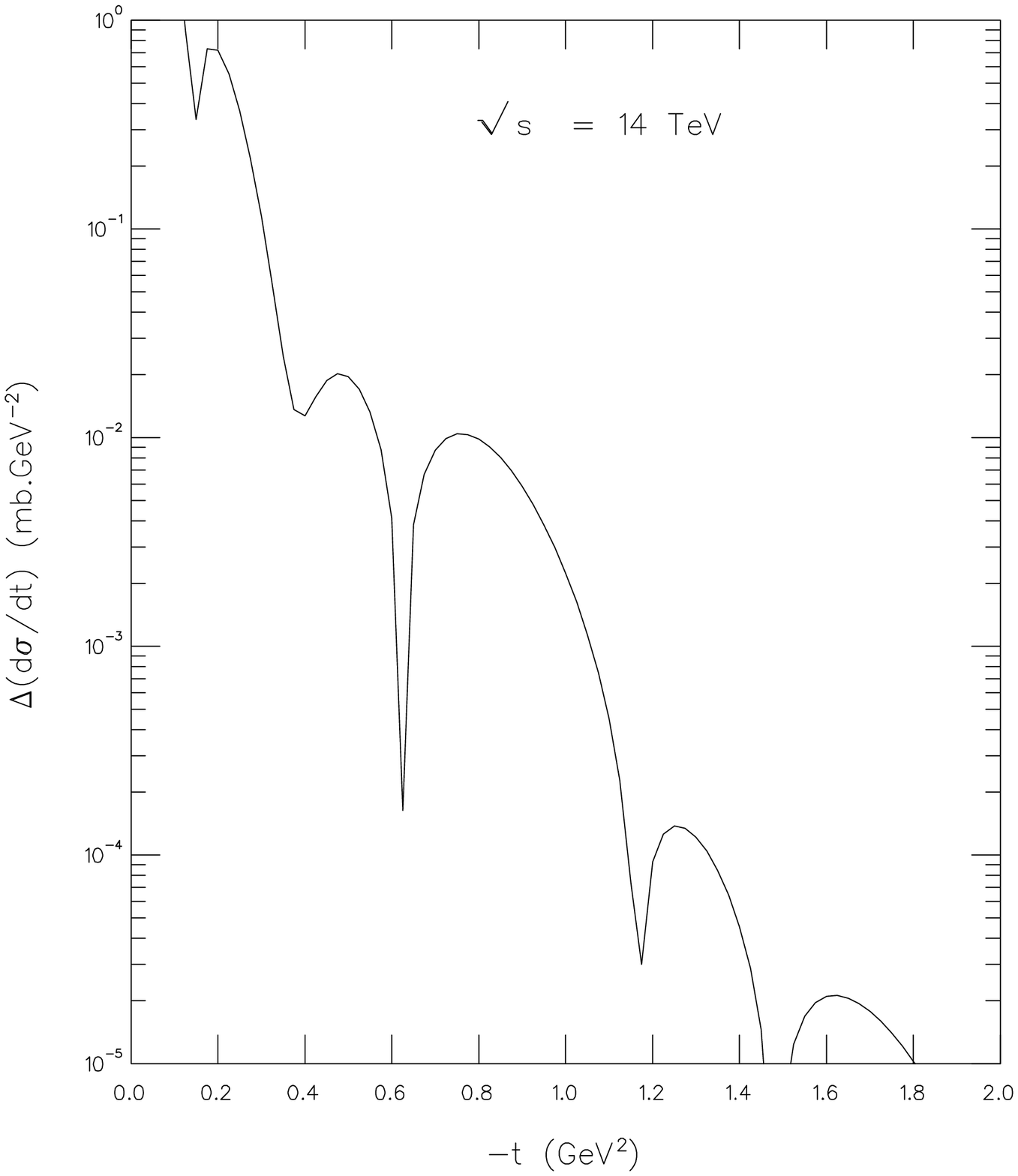}
\caption{Oscillations in the difference between the $pp$ and $\bar pp$ differential cross-sections $\Delta\left(
\dfrac{d\sigma}{dt}\right)(s,t)\equiv
\left\vert
\left(\dfrac{d\sigma}{dt}\right)^{\bar pp}(s,t)-
\left(\dfrac{d\sigma}{dt}\right)^{pp}(s,t)
\right\vert$}
\end{figure}

\end{document}